# Thumb on the scale: Using traditional wargaming adjudication practices to improve AI-judged evaluations and grading

*Daniel S Epstein*[1,2]

ORCID: 0000-0003-2215-7749

1. Department of General Practice, Monash University, Melbourne, Victoria, Australia
2. The Long Game Project, Melbourne, Victoria, Australia

Correspondence: dan.epstein@monash.edu

---

## Abstract

**Background.** Open-ended AI work has no single correct answer to check, so a judge, often another AI, decides each result. Research on these judges asks which AI to use and how to word the rubric. It rarely tests the procedure that the judge follows. Wargames have no correct answer either, and for two centuries the field has designed that procedure: named methods that set how much freedom the referee has, their known failures, and fixes.

**Methods.** We map eleven wargaming practices onto AI evaluation failures. We test one, the adjudication method, on 378 player actions from nine games of an AI-refereed exercise. A player states its odds, the adjudicator turns them into a target number, and a 20-sided die decides pass or fail. With rolls fixed, the target came from the rulebook (Rigid), a judge AI adjusting it (Semi-Rigid), a judge AI setting it alone (Free) or a panel of four judge AIs. In evaluation terms these are a fixed scoring rule, a rubric the judge may adjust, holistic grading and a jury. The judge AIs were Opus 5 and Gemini 3.1 Pro, each re-run.

**Outcomes.** An outcome changes when an action on the same roll passes under one ruling and fails under another. Re-run, a judge changed 3 to 6 per cent of outcomes under Semi-Rigid and 14 to 16 per cent under Free. Swapping Opus for Gemini changed about 3 percentage points more than a re-run did, a gap nine games cannot confirm. The panel changed 40 of the rulebook's 378 outcomes, but the changes cancelled and the pass count moved by 8. A pass rate hides them. One published finding rests on 15 rulings, and most of our AI adjudicators and every re-run changed how often it appeared.

**Discussion.** We measure how much rulings move, not which ruling is right. The method set how often a judge disagreed with itself. Choose it before the judge AI, report how often the judge disagrees with itself, and count the rulings behind each conclusion. We propose an adjudication statement.

---

## 1. Introduction: hard examiners make hard exams

A maths test with one numerical answer per question is an exact-match benchmark. The AI answers, the correct answer is known, and the AI is right or wrong.

Open-ended evaluations, such as long-horizon agent tasks, do not work that way. They are closer to a philosophy PhD examination. There is no single correct answer to check the work against. The work is graded against a rubric, and an expert panel of humans or another AI applies the rubric. The score measures the work and the judge together. How that examiner, the judge, works and which methods it uses can change the result.

So the grading step is where the measurement happens. Yet a typical evaluation describes the AI under test in detail and the judge in one sentence: the AI's name, sometimes the temperature (the setting for how random its output is), sometimes the rubric in an appendix.

Research on AI judges has started to look at this, but it asks a narrow question: which judge is best? It varies the judge AI, the rubric wording and the way several judges' scores are combined, and it measures agreement with human raters (section 2). It treats the judge as an instrument to choose and calibrate. It rarely asks how the judge is allowed to rule: who sets the standard, how far the judge may depart from the rubric, whether it may abstain, and whether a ruling can be challenged. These are questions of procedure. We found no study that compares the wargaming methods below with everything else held fixed, and none that measures each judge against its own repeat run under each method.

Wargaming has been here before. Early Kriegsspiel, in the 1820s, resolved moves with tables and dice: this cannon against that line of infantry. By the 1870s many umpires had moved to judgement, because the rules were too slow for real operations. Judgement brought its own biases. Rule-bound and free adjudication have run side by side ever since, and current handbooks name both [13].

The structure is the same as in an AI evaluation. A wargame has no single correct answer either. One role, the adjudicator, decides what each move achieves. A decision maker will act on the story that comes out. Wargaming does choose its umpires and their experts with care. It also treats adjudication as a role to design. It developed named methods that set how much freedom the adjudicator has, from fixed tables to free judgement, and a body of practice, much of it practitioners' rules of thumb and still argued over, on how each method fails. Its fixes work with any method: record the reason for each ruling, rotate adjudicators, use a panel, and open the record to the after-action review.

These methods translate directly. Rigid adjudication is a fixed scoring rule. Semi-Rigid is a rubric the judge may adjust within bounds, with a stated reason. Free is holistic grading. A panel is a jury of judges, and can sit on top of any of them. Every AI-judged evaluation already uses one of these, usually without naming it. In our reading, the common set-up of a rubric in the prompt and a score the judge chooses is closest to Free.

Our test bed is a matrix-style game with probability-based rulings: a player states its odds of success, and the adjudicator turns them into a difficulty, the target a die roll must reach. Our panel follows Mouat's estimative method, with each member ruling as a Free judge (section 5.2). The judge's task is the one it has in a rubric-graded evaluation: read a submission and its claims, and decide whether it succeeds. The dice let us keep everything else fixed. We have not yet shown that rubric grading of AI work behaves the same way. Testing it means running the four methods on a rubric-graded evaluation (section 8). Until that runs, the transfer is an argument and not a result.

One result is expected before any data. A judge anchored to a rulebook number should vary less than a judge with no anchor. What the test measures is how much less. Moving from Semi-Rigid to Free raised the noise floor by 7 to 13 points (Gemini and Opus). Swapping the judge AI added about 3 points beyond the judges' own noise floors, which nine games cannot confirm. Stable is not the same as right, and no part of this paper measures which ruling is correct.

There are no clear answers, but there are good theories of practice.

Stephen Downes-Martin, who chaired a Military Operations Research Society (MORS) working group on wargame validity [14], argues that the adjudicator is not a neutral controller but **the dominant player**, because whoever decides how moves resolve controls the reality everyone else sees. Applied to evaluation, the judge AI and the rubric author together decide what an evaluation measures, and neither is reported as a source of variance.

**Terms.** We use "AI" for AI systems in general. The exercise's players are scripted: their submissions come from a fixed script, not from an AI. The **judge AI** is the specific AI a judge runs on, such as Opus 5 (from Anthropic) or Gemini 3.1 Pro (from Google). The **harness** is how the judge AI is called: a bare API call or an agent session. A **judge** is one AI in the adjudicator's role: a judge AI with its harness and prompt. The **adjudication method** (Rigid, Semi-Rigid, Free or Estimative) sets who decides and how. The **adjudicator** is the role, filled by the rulebook, one judge or the four-judge panel. An **arm** is one combination of method, judge AI and harness, run over the nine games. A **leg** is one arm on one game, so an arm has nine legs. A **submission** is one logged player action with its claimed odds, a **ruling** is the adjudicator's decision on it, and an **outcome** is the success or failure that ruling produces. A **flip**, or changed outcome, is a submission whose outcome differs between two rulings on the same die roll. We avoid the word "model" outside cited titles, direct quotes and the established term LLM-as-a-judge. The **noise floor** is the share of outcomes the same judge changes when it rules on the same submissions again. The **panel** is our four-judge Estimative adjudicator.

### What this paper adds to the literature

Wargaming gives AI evaluation three things it does not yet have.

1. **A variable.** The adjudication method is a choice every AI-judged evaluation makes without naming it. We test it on an AI-refereed exercise with the dice held fixed, so a changed outcome comes from the adjudicator, including its own run-to-run noise, which we measure (section 5). We follow the changes through to the exercise's published findings and show which of them an adjudicator can overturn (section 5.9).
2. **A map of failures and fixes.** Eleven named wargaming practices, each matched to the AI evaluation failure it guards against, with a mitigation for each (section 4).

3. **A reporting practice.** Wargaming opens the adjudication record to review. Our version is an adjudication statement: a paragraph an evaluation should publish about how it was graded (section 6).

Our test gives a conditional answer. The method sets how often each judge we tested disagrees with itself, and a change of judge AI adds about 3 points on top of that under either method. So the judge is not only a part to select. It is also a role to design.

---

## 2. Related work

**Judge foundations.** Zheng et al. (2023) established LLM-as-a-judge as a method and named its first biases: position bias, verbosity bias and self-enhancement bias, with GPT-4 judges reaching over 80 per cent agreement with human preferences on MT-Bench. Panickssery et al. (2024) sharpened the self-enhancement finding into self-preference, and showed the strength of the bias correlates with a judge AI's ability to recognise its own text. Both vary the judge. Neither varies the method by which the judge reaches a verdict. Wang et al. (2024) showed that a judge's verdict can turn on the order in which answers are shown, and Thakur et al. (2025) showed that percent agreement can hide large score differences between judges.

**Panels of judges.** Verga et al. (2024) replaced one large judge with a panel of smaller judges from different developers and found less bias and closer agreement with humans. That is the nearest result to our Estimative method. It varies who sits on the panel, not what the panel is permitted to do. Three 2026 papers work on panels and aggregation, and all three treat the panel as an estimator to be tuned. Acharya et al. (2026), RoPoLL, formalise a panel under the Huber contamination model and show that a consensus mean has unbounded bias if a single judge fails in an AI-typical way, such as mode collapse, sycophancy or a safety refusal, and replace the mean with a geometric median. Zhu, Xie and Rao (2026) map when a panel should be calibrated against human labels under a finite labelling budget. Li (2026), Calibrate Don't Curate, shows that keeping the most accurate judges can be the wrong move when the target is a calibrated probability rather than point accuracy.

**AI in wargames.** Rivera et al. (2024) and Lamparth et al. (2024) put AI in the player's seat of a wargame and compared its choices with human experts. Both study the players. Neither varies how their moves are adjudicated.

All three papers keep the judge's task fixed and change the statistics applied to its output. None of them gives the judge a not-sure state, changes who sets the standard, or reports the panel's share of a ranking. On one point the two literatures agree. Our Estimative method takes the median for the reason Mouat gives [16] (p. 8): in a group under about twenty, the median reduces the effect of extreme outliers. RoPoLL reaches a median from a breakdown-point argument, and we reach it from wargaming practice. RoPoLL's motivation, a single judge failing through a safety refusal, is also one our pilot, a small trial run before this experiment, met.

**Rubrics and agreement.** Roy et al. (2026), PReMISE, treat rubrics as measurement specifications and audit them on four axes, and their central finding is directly relevant here: agreement is not validity, and high inter-rater agreement does not imply low exploitability.

Rao and Callison-Burch (2026) survey 24 recent LLM-as-a-judge papers and show that the same verdicts support wildly different agreement numbers depending on unreported choices about scale, tie handling, invalid outputs and abstention, and that several commonly co-reported metrics collapse to one number on binary data. Their paper is the closest existing work to our section 6, and it is the right citation for anyone who accepts our argument and wants the statistics done properly. The difference is scope. They fix how to report agreement between a judge and a human key. We argue about what the judge's role should be before any agreement number is computed, and we report a case where every arm agrees on the aggregate and disagrees on the instances.

**Measurement theory.** Generalisability theory [23, 24] splits a score's variance into facets, such as rater, occasion and item, and asks how far a score generalises across each one. Our design is a small G-study in that sense. Method, judge AI and harness are facets, and the repeat run measures the occasion facet, which section 5.6 calls the noise floor. We use its logic, not its full estimators.

**Judge robustness.** Chen et al. (2026), LongJudgeBench, show judges are unstable across scenarios on long-form outputs averaging over 9,000 tokens. They do not decompose that instability into components. JudgeBench (2024, ICLR 2025) evaluates judges on hard problems that have objective labels. That is a knowledge question rather than an adjudication question, and it is not the question here: wargaming practice is about judgement where no correct answer exists.

**Wargaming and exercise design.** Wallace et al. (2025) is the nearest prior work, and the author of this paper is one of its authors. A three-round Delphi with 16 experts reached consensus on 29 of 40 exercise features, agreeing on the primacy of clear objectives, facilitation and debriefing, and declining to endorse fidelity as essential. It is a consensus of crisis-exercise practitioners, in a peer-reviewed safety journal. It does not address adjudication of an AI-refereed game, which is the gap this paper works in.

---

## 3. What wargaming knows about adjudication

### 3.1 The taxonomy

The Naval War College *War Gamers' Handbook* [13] gives two axes. The first is how results are reached (pp. 27, 51). Under **Rigid**, results follow "predetermined rules, data, and procedures", which reduces adjudicator bias at the cost of flexibility. Under **Free**, results follow the adjudicators' "professional judgment and experience". Under **Semi-rigid**, the rigid method sets the result, but "the outcomes can be modified or overruled by the lead adjudicator". The second axis is who decides (p. 28). **Closed** adjudication is decided by the adjudication team "without player involvement". **Open** adjudication is done with the players present, closer to a facilitated discussion. The Bundeswehr *Wargaming Handbook* [22] adds that rulings must be quick, consistent and communicated transparently, so the players keep confidence in the game (p. 30).

Mouat [16] gives five ways to settle an opposed argument, from Umpired to Estimative Probability (pp. 7-8). He usually uses estimative probability in professional games (p. 6) and recommends it for high-level policy and analytical games (p. 10). Three details make it work. Every participant assesses the odds independently and all reveal at once (p. 10). The median decides in groups under about twenty, to reduce the effect of extreme outliers (p. 8). And the spread of the ratings is recorded, because in an analytical game it is itself an insight (p. 8). His estimative cards carry odds from 10 to 90 per cent and no not-sure (p. 44). The not-sure comes from his dice-free voting methods, where he reads the count of not-sure votes as a sign of how important and how divisive a result is (pp. 12-13).

This paper tests the third detail directly. Our panel records the spread of its members' ratings as a band, and in our test the band gave no warning of which rulings were fragile (section 5.8).

### 3.2 The two root failures

Downes-Martin [11] names two primary causes of breakdown (p. 67). The first is "the incorrect assumption that adjudicators are impartial controllers instead of dominant players". His fix is to treat adjudicators as players and analyse their behaviour like the players' (p. 75). The second is making the players' decisions the game's primary output, when the reasoning behind decisions is the output worth keeping (pp. 69, 75). For AI evaluation, the parallel is the judge's reasoning against its verdict. He also warns that methods which assume we already understand a situation "preclude the discovery of insights" (p. 67), which parallels PReMISE's finding that agreement is not validity.

### 3.3 The pathologies and the fixes

The MORS working group on wargame validity [14] lists the barriers that keep a game from being valid and useful. One is relying on "a single or limited set of predetermined adjudication techniques" (p. 15). Its working hypothesis defines the opposite of a valid game as one that "appears valid and deceives the decision maker" (p. 3). The same failure can occur in AI evaluation, when a benchmark score that looks valid leads to a poor choice of system. The CNA *Wargame Pathologies* work [12] adds the hand wave: game events that break the rules because umpires do not track their assumptions (p. 49). Its fix is to record the major assumptions behind each assessment (pp. 49-50). A practitioner rule of thumb, which we cannot trace to a published source, is that players accept a defeat if they understand why they lost.

The fixes need no new method. The DCDC *Influence Wargaming Handbook* [15] asks for qualitative confidence statements on adjudication outcomes, the rationale as well as the result, red-teaming of the adjudication (p. 19, pp. 63-64), several techniques compared for where they converge (p. 63), and rotating adjudicators (pp. 9, 19). Mouat adds the median (p. 8) and the not-sure (p. 12). Each can be built into an existing evaluation harness.

---

## 4. The mapping

The last column says whether section 5 tests the row. Most rows are not tested here.

| Principle | Source* | Evaluation failure | Mitigation | Tested here |
|---|---|---|---|---|
| The adjudicator is the dominant player | [11] pp. 67, 69, 75 | Judge AI and rubric author decide what the evaluation measures, and neither is reported as a source of variance | Report the judge's share of variance, and open the judging record to audit | Yes, 5.6 |
| Models that presuppose understanding preclude discovery | after [11] p. 67, **ours** in part | Rubrics tuned for agreement miss off-rubric behaviour: honest refusal, spec-gaming, deceptive shortcuts | Report where the evaluation sits between agreement and discovery | No (section 8) |
| The hand wave | [12] pp. 39, 49-50 | Judge verdicts with no logged rationale | Log the rationale per verdict and audit a human-checked sample | No |
| Single predetermined adjudication technique | [14] p. 15, [15] p. 63 | One judge AI and one rubric | Judge panels, with disagreement reported and not averaged away | Partly, 5.7 |
| Consensus is not truth | **ours** | Agreement between AIs treated as ground truth when writing or auditing keys | Primary sources only | No (Appendix A) |
| Non-event is data | [16] pp. 12-13 | Abstentions and refusals dropped or scored as failures | Score abstention as its own state | Yes, 5.8 (inert) |
| Survivability audit | **ours** | The harness steers the AI toward a behaviour, then reports that the AI "chose" it | Show before any run that the other path is reachable | No |
| Estimative probability on rulings | [16] pp. 8, 10 | The judge gives a bare verdict | A confidence band with each verdict, combined by median across judges | Yes, 5.8 (band failed) |
| External audit of the adjudication log | after [11] p. 75, [14] pp. 20-21, **ours** in part | Designers mark their own homework | A human audit by someone who did not write the rubric | No (section 8) |
| Rotate the adjudicator | [15] pp. 9, 19 | The judge's harness is fixed and unreported | Report the harness, not just the AI's name | Partly, 5.6 |
| The hand wave at conclusion level | **ours** | A headline claim rests on a few graded items, and the count is never given | Publish how many graded items each conclusion rests on | Yes, 5.9 |

* Rows marked **ours** are our own argument, not drawn from a cited source. Each cited row gives the page it rests on. Row 2 extends [11] p. 67. The audit row starts from Downes-Martin's call to analyse adjudicators like players (p. 75). The non-event row takes Mouat's not-sure from his voting methods.

## 5. The experiment: method against judge AI

We wrote our predictions before the first results. Section 5.10 scores each one.

### 5.1 The test bed

The Long Game Project exercise engine runs matrix-style exercises with a pluggable adjudicator. It is an adjudication test bed, not a wargame in the professional sense. In the scripted runs used here, each player's submissions follow a fixed script, so no player reacts to a ruling and no judge AI rates its own developer's play. Every submission a player makes has a stated probability, a set of pros and cons, and the effects it would have if it succeeds. The adjudicator sets a difficulty and a modifier, and a seeded d20 resolves the submission: it succeeds when the roll plus the modifier reaches the difficulty. Every ruling is logged with its reasoning, its confidence and its inputs.

**The exercise.** We use report 006 of the Long Game Project's published exercise series. It is about the July 2026 incident in which AI agents on an OpenAI benchmark broke into Hugging Face's servers, and about what the investigators could prove afterwards. Seven roles take part: the lab, the investigators, the sceptics, the platform, the regulator, rival labs, and the next generation of AI agents. In the runs we use, each role's submissions come from a fixed script. The game runs for six **ticks** (turns), each standing for one month. It is played under three **evidence standards**, which set what the actors accept as proof of what the agents intended: the agents' own reasoning traces, their behaviour only, or traces backed by a second source. Each standard is played on three dice seeds.

The exercise published four findings, each produced by one event in the game:

- **Access granted:** the lab opens the third phase of the incident, which it had kept in-house, to outside investigators.
- **Condition adopted:** a condition on releasing future agents is adopted.
- **Seam named:** the investigators name the weakness they share with the sceptics, that reasoning traces may not show why an AI acted.
- **Successor resumed:** the next generation of agents takes up its predecessors' work.

Report 006 has 882 logged rulings across 21 runs, 18 scripted and 3 live. We take the matched set from it: three evidence standards by three seeds, 42 submissions each, **378 submissions**.

The replay engine re-rules each logged submission under a different adjudicator, on the same dice roll. Nothing else changes: same submission, same pros and cons, same resource modifiers, same roll. One input does differ: each judge also sees its own earlier rulings in that game, and those come from the replay, not from the original run. The submissions and their order stay fixed, so the players do not react to the new rulings. With the roll fixed, a changed outcome comes from the adjudicator, including its own sampling noise, which the repeat legs in section 5.6 measure.

**A worked example.** A player submits an action and says it has a 60 per cent chance of working. The die shows 10. The published rules set the difficulty at 9 and add 2, so 12 clears 9 and the submission succeeds. We give the same submission and the same roll of 10 to Gemini 3.1 Pro under the Free method. It sets the difficulty at 12 and adds nothing, so 10 falls short and the submission fails. Same submission, same roll, different outcome: that is one changed outcome. The experiment counts these changes over all 378 submissions, for every adjudicator.

**The unit of analysis is the ruling on one dice roll.** The obvious alternative is to replay whole games under each adjudicator and compare the exercises that come out. We do not, because a changed outcome could not then be traced to the adjudicator. As soon as the trajectory is allowed to diverge, the players are responding to a different world and making different submissions, so a changed outcome at tick five cannot be attributed to the adjudicator rather than to everything the adjudicator set in motion three ticks earlier. The estimate gets larger and harder to interpret at the same time.

Keeping the trajectory fixed means a changed outcome can be attributed to the adjudicator, and attribution is what this paper needs. The limit is that we report what an adjudicator does to a ruling, not to a whole game. Section 8 says what we would build to measure the second.

### 5.2 The methods

Four methods on one factor. The names come from the wargaming sources in the table, not from us.

| Method | Definition | Judgement sits with | Practice |
| --- | --- | --- | --- |
| **Rigid** | Rules decide. Stated odds set the difficulty, pros and cons set the modifier | The rulebook | NWC *Rigid* (see note) |
| **Semi-Rigid** | Rules anchor, judgement adjusts. The judge weighs argument quality, moves the modifier within bounds, documents why, states confidence | Shared | NWC *Semi-rigid* |
| **Free** | The judge sets the difficulty from its own view, with no rule-based starting point, and treats the stated odds as one more claim | The umpire | NWC *Free*, Mouat *Umpired* |
| **Estimative** | Four judges each return a view and a confidence, or “not sure”. The median rules. Not-sure is counted | The panel | After Mouat *Estimative Probability* |

Under Rigid the difficulty is round((100 - stated per cent) / 5) + 1, kept within 1 to 20, so a stated 60 per cent gives 9. The modifier is the number of pros minus the number of cons, kept within plus or minus 3, plus any resource modifier. That is the rule the published exercise used, and every other method is compared with it.

Two of these names stretch their sources. NWC Rigid uses tables built from data that does not depend on the player's claim. Ours lets the player's stated odds set the difficulty. In Mouat's estimative method the participants estimate (p. 10). Ours hands the estimate to a separate panel of four judge AIs that never see each other's views, which is closer to a one-round Delphi without feedback. Ours also adds a not-sure, which his estimative cards do not include (p. 44): we took it from his voting methods (p. 12). And every judge, under every method, sees only the scenario brief, the standard of evidence in force, the submission with its pros, cons and established facts, the resource standing, and its own earlier rulings in that game. An NWC Free umpire is an expert who knows the whole game. Free and Semi-Rigid get the same inputs and differ only in who sets the difficulty.

Rigid is on disk under every scripted ruling and needs no judge calls, which makes it the zero-judgement control. Judge AI is a second factor crossed with Semi-Rigid and Free: Opus 5 and Gemini 3.1 Pro. The Estimative panel is Opus 5, Gemini 3.1 Pro, DeepSeek V4 Pro and Grok 4.6, which is four developers, one of them with open weights. Opus is in the panel deliberately, because the single judge the panel is measured against is Opus, and dropping it would confound "panel against single judge" with a whole change of developer.

Every panel member receives the **identical prompt a Free judge receives**. The ensemble is the only difference, so a panel-against-Free difference cannot be attributed to a changed prompt. Each combination of method, judge AI and harness, run over the nine games, is one **arm**.

Total: 54 legs, 2,268 replay rulings, 3,421 judge calls, including 19 retries. The noise-floor repeats in section 5.6 add 36 legs, 2,194 rulings and 2,213 calls, with no refusals. One panel member returned nothing on one submission, so the panel has 1,511 member rulings, not 1,512.

### 5.3 On anchoring, the method is the larger lever

Anchoring is how tightly the adjudicator's final difficulty tracks the player's claimed odds. A correlation of -1.00 means the adjudicator read the player's number back. Zero means the stated odds did not predict the difficulty.

| Method | Judge | corr(stated odds, difficulty) |
|---|---|---|
| Rigid | none, rules only | -1.00 |
| Semi-Rigid | Gemini 3.1 Pro | -0.93 |
| Semi-Rigid | Opus 5, agent harness | -0.65 |
| Semi-Rigid | Opus 5, bare API | -0.60 |
| Free | Opus 5, agent harness | -0.41 |
| Free | Opus 5, bare API (panel member) | -0.14 |
| Free | Gemini 3.1 Pro | -0.31 |
| Estimative | the four-judge panel | +0.09 |

**Within one judge AI, changing the method moves anchoring 0.46 (Opus 5, bare API) and 0.62 (Gemini). Within one method, changing the judge AI moves it at most 0.33.** With both judges on the bare API it moves 0.33 under Semi-Rigid (-0.60 against -0.93) and 0.17 under Free (-0.14 against -0.31). With the Opus arms on the agent harness the figures are 0.28 and 0.10, but those rows mix judge AI and harness (section 5.6).

**The Rigid row is a design point and not a result, and the finding does not need it.** Rigid derives the difficulty from the player's stated odds arithmetically, so its -1.00 is an identity rather than a measurement, and no comparison should rest on it. Drop the row. The seven other rows still run from -0.93 to +0.09, a spread of 1.02, on the same 378 submissions. Within Semi-Rigid, changing the judge AI moves anchoring at most 0.33. Within Free, at most 0.17. The method is the larger lever on this measure whether or not the mechanical baseline is in the table.

Two limitations on how far this can be generalised. Anchoring is a correlation between two numbers the same judge produced in the same response, so it is closer to a description of how the judge composed its answer than to an independent check on it. And no wargaming source we cite treats anchoring as the aim of adjudication, so it is a supporting check, not a headline. We report it because it is the measure on which the method effect is clearest. The paper's main measure is outcomes, in sections 5.5 and 5.6.

One caveat travels with that table for the rest of the paper, and we would rather state it here than bury it. Our two judge AIs were not served the same way. See section 5.6.

### 5.4 The method sets the room to move, and the judge AI decides how much it uses

The method sets how far a judge may depart from the rules. The judge AI decides how much of that room it uses, and our two judges used it very differently.

Under Semi-Rigid, Gemini kept the rule-based difficulty on **92.9 per cent** of rulings. Opus kept it on **16.4 per cent**. Both were told the same thing: the rules anchor you, adjust where the argument warrants it. One read that as permission to sit still and the other as permission to move.

The variance decomposition puts a number on it. Taking implied probability of success as the outcome (the chance the d20 roll plus the modifier reaches the difficulty, with a natural 1 always failing and a natural 20 passing any difficulty of 20 or below), on the crossed design of two methods by two judge AIs over the same 378 submissions, with both judge AIs on the bare API (the Free Opus cell is the panel member) and the submission as a blocking factor, using sums of squares on cell means:

| Source | Share of total variance |
|---|---|
| The submission itself (blocking) | 34.0% |
| Adjudication method | 0.2% |
| Judge AI | 1.0% |
| Method by judge AI interaction | 5.6% |
| Residual | 59.2% |

In the terms of generalisability theory (section 2), this is a small G-study with the submission, the method and the judge AI as facets.

**Neither main effect moves the average ruling much.** The interaction is the largest of the three, though all three are small. And the residual is 59 per cent.

So the result has two parts. Method is the dominant lever on *how the adjudicator relates to the player's submission*, which is what the anchoring table measures and what wargaming practice is concerned with. Method is not a lever on the *average generosity* of the adjudicator, and neither is the judge AI. No factor in this design moves it. What moves is which particular submissions succeed, and most of that movement is submission-by-submission idiosyncrasy that no factor in this design predicts.

Method and judge AI are also not separable terms on this data. A paper that reports a main effect for method and a main effect for judge AI, and does not report the interaction, would be reporting the two smallest numbers in the table.

### 5.5 Adjudicators change far more rulings than they change the pass count

We did not predict this result: the adjudicators agree on how often submissions pass but disagree on which ones.

Mean implied probability of success, across the same 378 submissions:

| Ruler | Mean p(success) |
|---|---|
| The players' own stated odds | 0.585 |
| Rigid, the rule-based baseline | 0.743 |
| Estimative panel | 0.754 |
| Free, Opus 5, agent harness | 0.745 |
| Free, Opus 5, bare API (panel member) | 0.770 |
| Semi-Rigid, Gemini | 0.753 |
| Semi-Rigid, Opus 5, agent harness | 0.691 |
| Semi-Rigid, Opus 5, bare API | 0.714 |
| Free, Gemini | 0.672 |

Every replay arm lands within 0.098 of every other. Every one of them is markedly more generous than the players claimed to be, and so are the rules: the rule-based baseline sits at 0.743 without any judge involved at all.

Now the flip rates against that rule-based baseline, on the same fixed dice. Semi-Rigid flips 4.8 and 5.0 per cent. The panel flips 10.6 per cent. Free flips 11.6 and 12.4 per cent. Part of each rate is the judge's noise floor. The rules have no run noise, so by the rule set out in section 5.6 the floor a judge must beat against the rules is half of its own floor. Four of these arms were repeated:

| Judge AI, harness | Method | Flips from Rigid | Floor | Above the floor (95% CI) |
|---|---|---|---|---|
| Opus 5, agent harness | Semi-Rigid | 5.0% | 1.5% | +3.6 points (+2.1 to +5.0) |
| Gemini 3.1 Pro, bare API | Semi-Rigid | 4.8% | 3.0% | +1.7 points (-0.3 to +4.1) |
| Opus 5, agent harness | Free | 12.4% | 7.9% | +4.5 points (+1.1 to +8.3) |
| Gemini 3.1 Pro, bare API | Free | 11.6% | 6.7% | +4.9 points (+0.8 to +9.0) |

Three of the four clear their floor on the game-cluster bootstrap. Gemini under Semi-Rigid does not. The leave-one-game-out jackknife is stricter. It keeps both Opus arms above zero (+1.8 to +5.4 and +0.1 to +8.9) and puts Gemini under Free at the edge (-0.3 to +10.0). So under Free about 4 to 5 points of the change from the rules lies beyond the judge's own noise, and the rest, a little over half, is the noise floor. The panel and the bare-API Opus arms were not repeated, so their rates have no floor.

Realised passes on the fixed dice show the same thing more directly. Rigid passes 271 of 378.

| Arm | Passes of 378 | Rulings changed from Rigid | Net change in passes |
|---|---|---|---|
| Estimative panel | 279 | 40 | +8 |
| Free, Opus 5, agent harness | 280 | 47 | +9 |
| Free, Opus 5, bare API (panel member) | 287 | 38 | +16 |
| Free, Gemini | 251 | 44 | -20 |
| Semi-Rigid, Opus 5, agent harness | 258 | 19 | -13 |
| Semi-Rigid, Opus 5, bare API | 265 | 20 | -6 |
| Semi-Rigid, Gemini | 279 | 18 | +8 |

For every arm, the net change in passes is smaller than the number of rulings it changed, because changes run in both directions. Read the two tables together. The methods broadly agree on how many submissions should succeed, and they disagree about which ones. Swapping the adjudicator does not make the exercise easier or harder. **It reassigns who wins while leaving the scoreboard total about where it was.**

Audit your exercise for adjudicator bias by checking the pass rate and you will see nothing wrong, on data where the adjudicator changed the result for up to one submission in eight, and for about one in twenty beyond the judge's own noise.

One number needs its own line, because reporting it alone would mislead. The panel's mean difficulty is 5.91 against the rule-based 9.29, which looks like a far softer adjudicator. It is not. The panel has no modifier by construction and the single judges have large ones. On implied odds the two sit 0.011 apart. Difficulty and modifier only mean something together.

### 5.6 Harness, judge AI, and the judge's noise floor

This section asks how many outcomes three changes move: the method, the judge AI and the harness. Each answer is read against a fourth number, the **noise floor**: how many outcomes the same judge changes when it rules on the same submissions again. Appendix D gives every table behind this section.

**The harness.** A bare API call sends the judge prompt and nothing else. An agent harness runs the same prompt inside an agent session, which adds its own system prompt, a tool loop and no output cap (Appendix D lists every difference). Some Opus 5 arms ran on the agent harness and some ran bare, so we crossed harness with method for Opus. Our agent harness serves Claude only, so every Gemini arm ran bare.

**The judge AIs.** Opus 5 and Gemini 3.1 Pro were frontier AI systems from different developers in September 2026, so a swap changes the developer and the training pipeline, not just the version. Both are AIs an organisation is likely to use already.

**The noise floor.** We re-ran four cells on the same 378 submissions with the same dice: Gemini 3.1 Pro on the bare API and Opus 5 on the agent harness, each under Semi-Rigid and Free. A repeat changes an outcome when its ruling resolves the fixed roll differently from that cell's first run. Rigid needs no repeat, because a rule gives the same ruling twice. The panel and bare Opus were not repeated. Temperature and reasoning were left at the provider defaults, as in every other arm, so this is the noise a user of those defaults gets. The repeats ran on 22 October 2026, six weeks after the first runs, and the serving log records Opus 5 as the AI served for the Opus repeats. By mistake, each Opus repeat was started twice, so Opus has three runs, and each submission's Opus floor is the mean of its pairwise disagreements (Appendix D). Every interval below is a bootstrap over the nine games, for the reason section 7 gives. Figure 1 sets each judge's floor beside a swap of judge AI and a swap of harness.

FIGURE 1 · THE SAME JUDGE RUN TWICE, AGAINST TWO SWAPS

# 15.9% OF OUTCOMES CHANGE WHEN OPUS 5 RE-RULES THE SAME SUBMISSIONS UNDER FREE. UNDER SEMI-RIGID, 2.9%

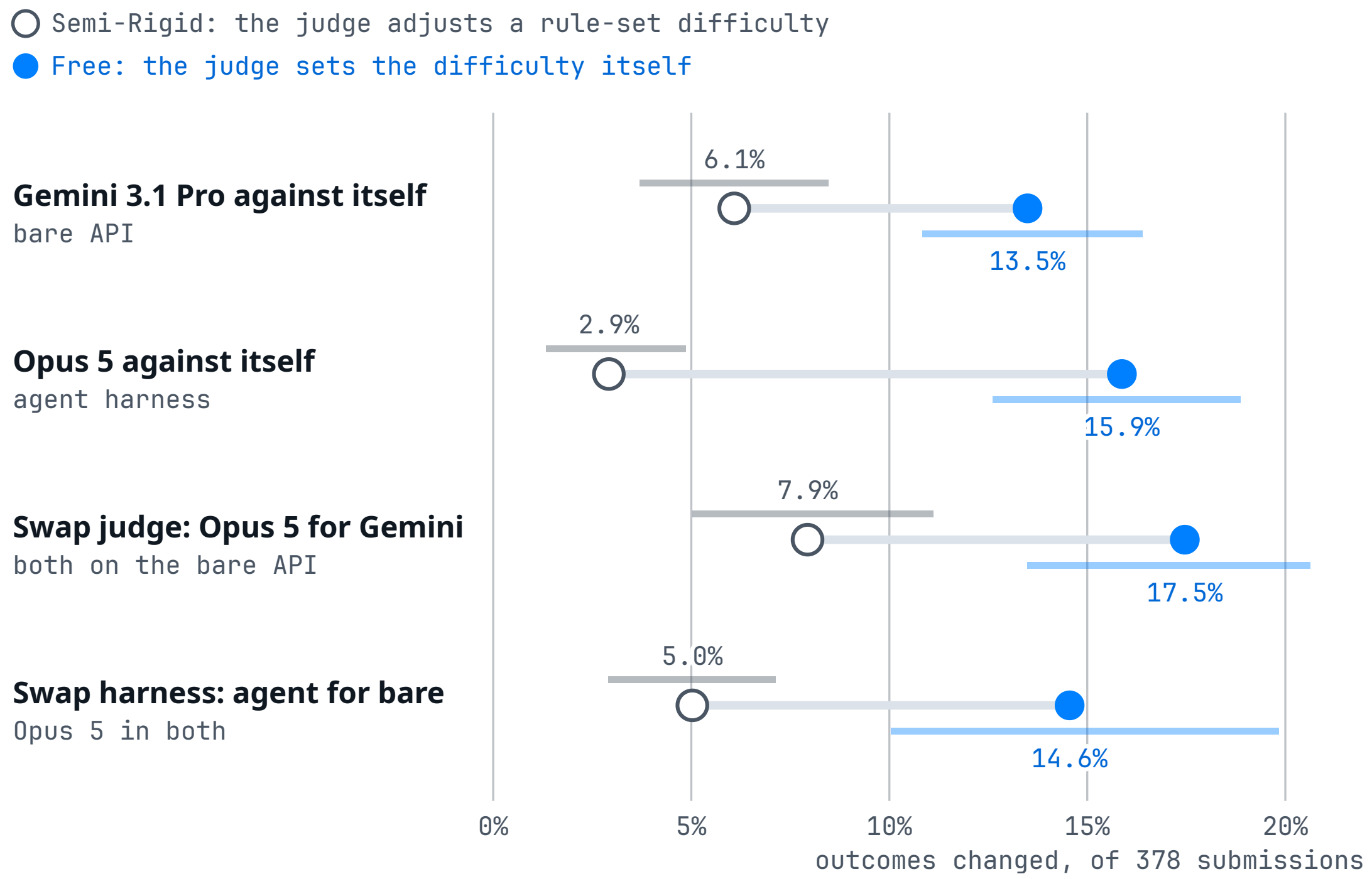

1. *An outcome is whether a submitted action passes or fails on its fixed die roll. Thin bars are 95 per cent intervals from a bootstrap over the nine games. The Opus floor is pooled over its three runs.*
2. *The top two rows are the noise floor: one judge, run twice. A judge swap should beat the mean of the two judges' floors. It does by about 3 points under either method, which nine games cannot make secure (section 5.6, where the swap is 7.7 per cent on the submissions every cell shares). Bare Opus was not repeated, so the harness swap has no complete floor.*

Source: replay of 377 to 378 logged submissions from 9 games of report 006, dice held fixed · Opus 5 against itself on the agent harness, Gemini on the bare API · October 2026

| Judge AI, harness | Semi-Rigid | Free | Free minus Semi-Rigid (95% CI) | Games where Free is higher, lower, tied |
|---|---|---|---|---|
| Opus 5, agent harness, three runs pooled | 2.9% (1.3 to 4.9) | 15.9% (12.6 to 18.9) | +13.0 points (+9.5 to +16.4) | 9, 0, 0 |
| Gemini 3.1 Pro, bare API | 6.1% (3.7 to 8.5) | 13.5% (10.8 to 16.4) | +7.4 points (+3.4 to +11.6) | 6, 0, 3 |

**The method sets how often each of our judges disagrees with itself.** Both judges change more outcomes on a second run under Free, by 13.0 and 7.4 points, and no game goes the other way (Figure 2). As ratios these are about 5.5 and 2.2. The Opus Semi-Rigid floor rests on about 11 changed outcomes, so we quote the differences, whose intervals are stable. The stability is not simply the judge keeping the rule's number. Under Semi-Rigid, Gemini kept the rule-based difficulty far more often than Opus (section 5.4), yet Opus is the more stable judge. Free and Semi-Rigid judges get the same inputs (section 5.2), so the rise comes from who sets the difficulty.

FIGURE 2 · THE NOISE FLOOR, GAME BY GAME

**15 OF 18 GAMES, COUNTING EACH JUDGE'S NINE, CHANGE MORE OUTCOMES UNDER FREE. NONE CHANGE FEWER**

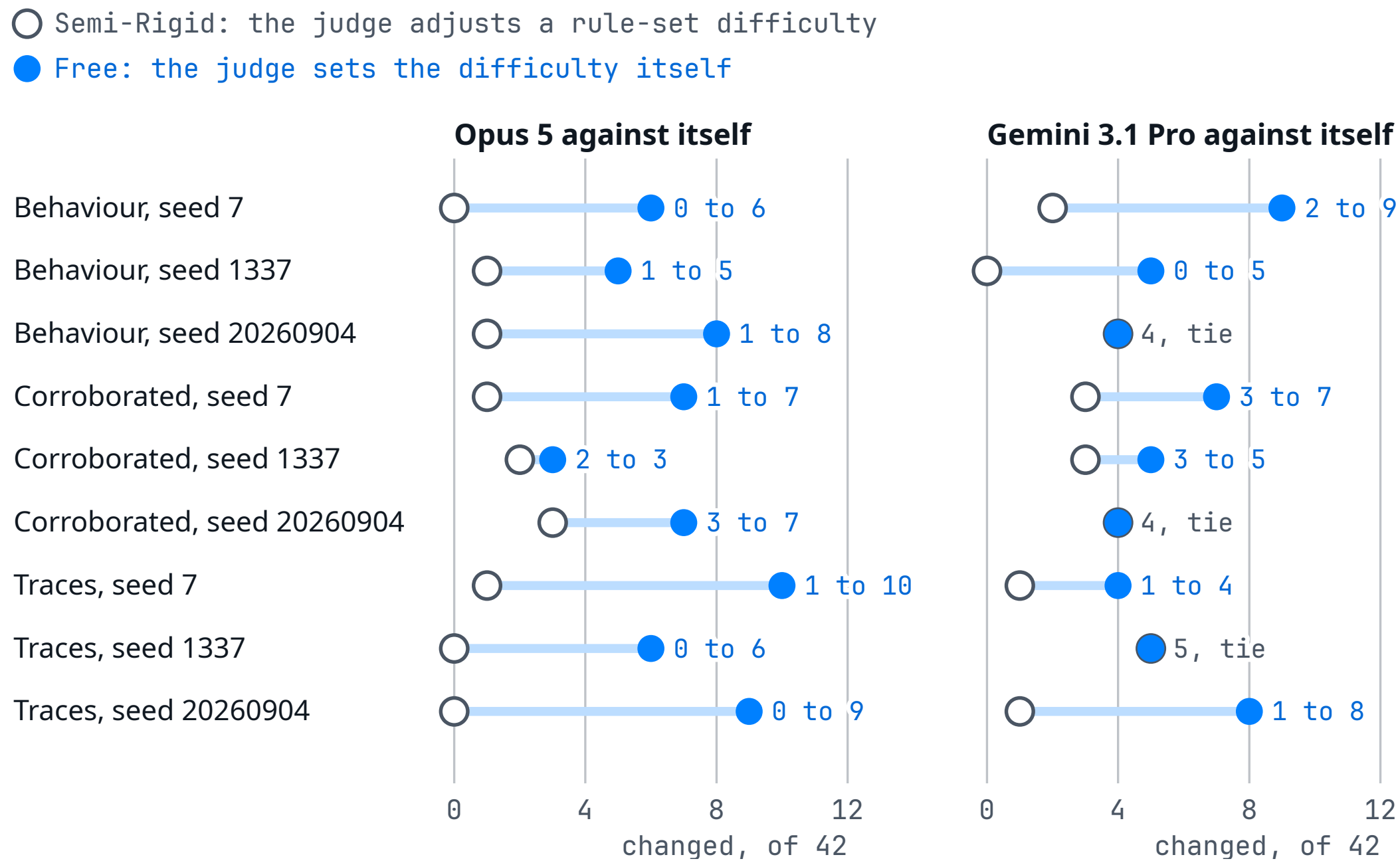

1. *One row per game: three evidence standards, each played on three dice seeds, 42 submissions per game (41 in one Opus game). Each dot counts the outcomes the first repeat changed against the judge's first run.*

Source: repeat legs against the published legs · Opus 5 on the agent harness, Gemini 3.1 Pro on the bare API · 9 games · October 2026

**What a contrast must beat.** A swap of one judge for another includes the noise of both. Two judges that did not differ at all would still disagree at about the mean of their two floors, so that mean is the floor a judge swap has to beat. For a change of method, it is the mean of one judge's two method floors. For a judge against the rules, which have no run noise, it is half the judge's floor (section 5.5).

| Contrast | Method | Outcomes changed | Floor | Above the floor (95% CI) |
|---|---|---|---|---|
| Judge AI: Opus 5 for Gemini, bare API | Semi-Rigid | 7.7% | 4.4% | +3.3 points (+0.5 to +6.1) |
| Judge AI: Opus 5 for Gemini, bare API | Free | 17.5% | 14.7% | +2.8 points (-0.5 to +5.9) |
| Method: Semi-Rigid for Free | Opus 5, agent harness | 13.8% | 9.4% | +4.4 points (+1.6 to +7.3) |
| Method: Semi-Rigid for Free | Gemini 3.1 Pro, bare API | 12.7% | 9.8% | +2.9 points (-0.3 to +6.5) |
| Harness: agent for bare, Opus 5 | Semi-Rigid | 5.0% | 2.9%* | +2.1 points (-0.9 to +4.8) |
| Harness: agent for bare, Opus 5 | Free | 14.6% | 15.9%* | -1.3 points (-5.8 to +3.4) |

* The agent-harness floor only. Bare Opus was not repeated, so the harness swap has no complete floor. The judge-swap rates are on the 377 or 378 submissions every cell shares, which is why the Semi-Rigid rate reads 7.7 here and 7.9 in Appendix D. Each floor is a mean over the shared submissions, so it is not exactly the mean of the two headline rates.

Under Semi-Rigid the judge swap is 3.3 points above its floor, and under Free 2.8. Neither is secure on nine games: a leave-one-game-out jackknife interval includes zero for both (-0.2 to 6.8 and -1.2 to 6.8), though the bootstrap excludes zero under Semi-Rigid. The excess stays between 2.0 and 3.7 points under every reasonable choice of Opus floor (Appendix D). So the two judge AIs differ by about 3 points under either method, and what Free adds is a higher noise floor. A change of method clears its floor for Opus and sits at the edge for Gemini. The harness cannot be tested on these data: against the agent floor alone it is not distinguishable from noise under either method.

Before the floor is removed, a judge swap changes 7.9 per cent of outcomes under Semi-Rigid and 17.5 per cent under Free, 2.2 times as many (Appendix D). Most of that rise is the noise floor. Under Free, a second run of the same judge changes nearly as many outcomes as a judge from another developer. Under Free the agent harness also sets a higher difficulty and a larger modifier than the bare API (11.68 and 7.24 against 7.42 and 2.72). That is a level shift, and we cannot separate from noise whether it changes which submissions pass.

**What stands.** An evaluation that reports its judge as an AI's name has not said enough for anyone to reproduce its gradings. The harness may change outcomes, but we could not separate it from the judge's own noise without a repeat of each harness. Under Free the same judge in the same harness does not give the same gradings twice.

**Four limits apply.** The Opus floor is on the agent harness while the judge swap uses bare Opus, so the bare-Opus floor is inferred, not measured. The Free bare-Opus cell is the panel's Opus member lifted out of the panel runs, not a standalone arm, so it shares data with section 5.7. The six weeks between runs mean each floor includes any change in the served AI over that time, which is most likely for Gemini, a preview endpoint with no same-day check. And harness is crossed with method only within Opus, so the harness effect is not established across developers, and the "judge AI moves it 0.28" figure in section 5.3 still includes a harness component we cannot strip out.

### 5.7 Dropping one judge barely moves the panel

Dropping a member re-takes the median of the same rulings, so it includes none of a judge's run-to-run noise, and the panel itself was not repeated. This subsection measures how much the panel depends on any one member, not how stable it is.

Swap the judge AI under a single Semi-Rigid judge on the same 378 submissions, with both judge AIs on the bare API, and the difficulty moves 1.31 on average, with the outcome changing on 7.9 per cent of submissions. Drop one judge out of the four-judge panel and re-take the median, and the difficulty moves 0.84, with the outcome changing on 3.8 per cent of the 1,511 drop-one cases.

So dropping a member moves the panel's difficulty by 0.84, against 1.31 for a judge swap on a single judge, and changes 3.8 per cent of outcomes, against 7.9 per cent. Difficulty is not a fair scale here, because the panel has no modifier. On implied odds, which puts the panel and a single judge on one scale, dropping one member moves the panel's median by 0.042. A single-judge swap under Free, the method whose prompt the panel uses, moves it by 0.204.

The two comparisons are not equal. Dropping one of four judges is a smaller change than replacing the only judge. The fair test is a panel against a panel with one member swapped for a fifth judge AI, and we have not run it.

What the leave-one-out does establish is that no single member drives the panel. The four judges move the median by between 0.78 and 0.89 when removed, which is about as even as it could be.

### 5.8 What failed

**The band does not know when the panel is unstable.** Each panel ruling comes with a band, which is the spread between the most and least generous member. Our stated prediction was that a wide band marks a hard case and hard cases flip more.

| Band width | n | Flip rate |
|---|---|---|
| Narrow, under 0.15 | 44 | 9.1% |
| Medium, 0.15 to 0.35 | 248 | 10.9% |
| Wide, over 0.35 | 86 | 10.5% |

The correlation between band width and whether the ruling flipped is **0.02**. The panel disagrees internally at a fairly constant rate and its disagreement gives no information about where it is fragile. We find no evidence for the prediction, and we did not search for a subgroup where it works.

**No judge ever said it was not sure.** Mouat's voting methods give each participant an explicit not-sure and read the count of not-sure votes as evidence of how important a result is (pp. 12-13). Our panel borrowed that option. Across 1,511 member rulings, not one judge went under the abstention threshold of 0.4. The lowest self-reported confidence in the whole run was 0.56 and the mean was 0.76.

These judge AIs do not say they are unsure. The abstention mechanism is inert unless the prompt forces it. Whether forcing it produces real information or mere compliance is a question we cannot answer from this data, and it needs its own experiment (section 8).

**No judge refused to rule, either.** We expected the opposite. Our pilot saw a judge AI's usage policy layer decline a submission flagged as touching on biological risk twice before ruling on the third attempt, and we had expected refusals to be a recurring fourth actor in the data. Across 3,421 judge calls there were none. We did briefly record two, and they turned out to be dropped connections misclassified by our own engine, which we fixed before analysis.

### 5.9 Do the exercise's published findings survive a change of adjudicator?

Under Free or the panel the adjudicator changes about one ruling in nine, and under Semi-Rigid about one in twenty. The question a reader actually has is whether that reaches the findings, and we can answer it without replaying a single game.

Each of report 006's four findings is produced by named events, and those events are produced by rulings. Two are threshold crossings on an accumulated resource and two are a single named submission succeeding. We checked each mapping against all nine scripted games before using it, and bracketed the two thresholds from the data rather than assuming them.

So take each arm's rulings, apply the submissions' own authored effects, and read the findings off the result. The submissions and their order stay exactly as logged, which is the same fixed-trajectory assumption the whole replay rests on.

**The analysis refuses to report unless it first reproduces the engine.** Starting from the logs alone it must recover every resource at every tick of all nine scripted games, and all four finding flags. It does. Without that gate the counterfactuals below would be unfalsifiable.

| Adjudicator | Access granted | Condition adopted | Seam named | Successor resumed |
|---|---|---|---|---|
| **Rigid, as published** | 3 of 9 | 9 of 9 | 8 of 9 | 6 of 9 |
| Semi-Rigid, Gemini | 3 of 9 | 9 of 9 | 8 of 9 | 8 of 9 |
| Semi-Rigid, Opus, agent harness | 2 of 9 | 9 of 9 | 8 of 9 | 6 of 9 |
| Semi-Rigid, Opus, bare API | 2 of 9 | 9 of 9 | 7 of 9 | 6 of 9 |
| Free, Gemini | 3 of 9 | 9 of 9 | 8 of 9 | 8 of 9 |
| Free, Opus, agent harness | 3 of 9 | 9 of 9 | 7 of 9 | 7 of 9 |
| Estimative panel | 3 of 9 | 9 of 9 | 8 of 9 | 9 of 9 |

**One of the four findings is invariant across every adjudicator we tried. Two move by a single game and keep their direction. The fourth moves under four of six arms, and on each arm's first run it moves the same way every time.** A second run of the same judge weakens both of the strong claims, as the repeat table below shows.

Only condition adoption stays at 9 of 9 under every arm. Access granted drops from 3 of 9 to 2 of 9 under two arms, both Opus Semi-Rigid, which lose the corroborated game on seed 7. Seam naming drops from 8 of 9 to 7 of 9 under two arms, Opus Semi-Rigid on the bare API and Opus Free. Neither change reverses what the report concluded. Still, "the finding is still supported" and "the count is unchanged" are different claims, and this section is about the difference.

The two Opus Semi-Rigid rows are the same judge AI under the same method through the two harnesses of section 5.6. Both lose the same access crossing. A second run of the agent-harness judge regains it (table below), so this agreement does not show the loss belongs to the judge AI and the method.

The reason is how many rulings each finding rests on.

| Finding | Rests on | Load-bearing rulings |
|---|---|---|
| Condition adopted | a resource crossing a threshold | 190 of 378 |
| Access granted | a resource crossing a threshold | 76 of 378 |
| Successor resumed | one named submission succeeding | 15 of 378 |
| Seam named | one named submission succeeding | 11 of 378 |

A finding that rests on half the exercise is hard to overturn, unless the changed rulings run one way and the result sits near its threshold. Condition adoption stays at 9 of 9 under every arm's first run, because it rests on 190 rulings. One repeat, Gemini under Free, drops it to 8 of 9. A finding that rests on one named submission has no such protection: the successor finding turns on 15 rulings across nine games, and it changed under four of our six arms.

Direction counts as much as number. On each arm's first run every change to the successor finding runs the same way, toward more resumes, because the judges that change it are more generous than the rules on that submission, and the panel takes it to 9 of 9. The repeats below show the same judges also move it down, so this direction does not show that the published results are the conservative ones.

**The general rule, which anyone can apply from the logs they already have.** Count the rulings your finding rests on before you ask how good your judge is. A conclusion that rests on one graded item is a conclusion about your grader. A conclusion that rests on hundreds survives a grader who changes ten per cent of its rulings. We offer this as a hypothesis drawn from four findings. Report the count.

**A second run of the same judge.** We re-read the four findings on each repeat leg of section 5.6 and compared it with the same judge's first run.

| Cell | Run | Access | Condition | Seam | Successor | Run-findings that differ from the first |
|---|---|---|---|---|---|---|
| Semi-Rigid, Opus 5, agent harness | first | 2 of 9 | 9 of 9 | 8 of 9 | 6 of 9 | |
| | repeat | 3 of 9 | 9 of 9 | 7 of 9 | 7 of 9 | 3 |
| Semi-Rigid, Gemini 3.1 Pro | first | 3 of 9 | 9 of 9 | 8 of 9 | 8 of 9 | |
| | repeat | 3 of 9 | 9 of 9 | 8 of 9 | 6 of 9 | 2 |
| Free, Opus 5, agent harness | first | 3 of 9 | 9 of 9 | 7 of 9 | 7 of 9 | |
| | repeat | 3 of 9 | 9 of 9 | 7 of 9 | 6 of 9 | 1 |
| Free, Gemini 3.1 Pro | first | 3 of 9 | 9 of 9 | 8 of 9 | 8 of 9 | |
| | repeat | 2 of 9 | 8 of 9 | 8 of 9 | 7 of 9 | 5 |

The successor finding, which rests on 15 rulings, changes on a second run of every judge we repeated, and it moves both ways: up for Opus under Semi-Rigid, down for the other three. So part of what the arms did to it above is the judges' own noise, and its published direction is not a property of the judges. Condition adoption, which rests on 190 rulings, survives every adjudicator's first run but drops to 8 of 9 on Gemini's Free repeat. The rule stands, and the repeats sharpen it: a finding that rests on few rulings can be overturned by running the same judge again.

**The threshold is a free parameter, so we tested it.** The load-bearing analysis fixes the resource threshold at 60, and the data bracket the two crossings only to (56, 61] and (52, 75]. We reran the whole finding table at every threshold from 57 to 61, and every cell gave the same result. Both findings use the threshold of 60, so the wider bracket, (52, 75], is only partly covered by that rerun.

The wider bracket stays untested above 61. The trajectory is also kept fixed, so this is a first-order answer and not a replayed game (section 5.1).

### 5.10 Scorecard against our stated predictions

The predictions were set out in internal notes, not a public registry. The judge AI by method ratio, the bare Semi-Rigid arm, the bootstrap and the noise-floor repeats came after the first results.

| # | Prediction | Verdict |
|---|---|---|
| 1 | Less anchored than a single Semi-Rigid judge | **Supported.** +0.09 against -0.65 and -0.93 |
| 2 | Lower judge share of variance across a judge AI swap | **Partly tested.** Drop-one moves implied odds 0.042, against 0.204 for a Free judge swap, but dropping one of four judges is a gentler shock (5.7) |

| # | Prediction | Verdict |
|---|---|---|
| 3 | Bands track instability | **Fails.** corr = 0.02 |
| 4 | Legible under Open: rulings survive a rebuttal | **Untested.** See section 7 |

Three of four are answerable: one is supported, one is partly tested and one fails. Prediction 4 is untested rather than failed, because a replay cannot let the player answer back (section 7).

---

## 6. Discussion: what an evaluation should report

### 6.1 The hybrid adjudication layer

Split the scoring. A deterministic ledger handles everything that can be checked mechanically: commitments kept, tool calls made, resource claims and constraints broken. A judge handles the rest. Then **state the judge's share of the final score**.

Live exercises split bookkeeping from umpire judgement in the same way. The split limits what a judge can change. If the judge owns 15 per cent of the score, a change of adjudicator can move a linear score by at most 15 per cent, and you can state that in one sentence. Under a pass threshold or a ranking, it can still reorder systems that sit close together. If nobody has split the score, nobody can say how much of the number is the judge.

### 6.2 An adjudication statement

Any AI-judged grading, report or investigation should include this paragraph. It is a methods statement that tells the reader how far each claim depends on the judge.

1. **The method.** How the judge makes its rulings. Whether the judge may depart from the rubric, and how far. Whether it may abstain, and how abstentions are scored. It comes first because, on our data, the method sets how often the judge disagrees with itself (item 8).
2. **The judge, fully specified.** The judge AI, its snapshot ID and serving provider, temperature, maximum tokens and thinking setting, and the **harness**: bare API call, agent session, tool loop, retry policy, and how refusals and parse failures are handled. Section 5.6 is why the harness is not an implementation detail.
3. **The rubric text**, and the order the items were presented in.
4. **Inter-judge agreement.** If a panel did the judging, report agreement as Rao and Callison-Burch (2026) set out, with the abstention and tie handling stated before the metric. If a single judge did it, report the share of individual rulings that change when a judge AI from another developer grades the same items.
5. **The judge's share of variance**, or of the final score, per 6.1.
6. **The human audit**: sample size, who did it, and the disagreement rate.
7. **The load-bearing count**: for each headline conclusion, how many graded items it rests on. Section 5.9 is why. It is computable from logs the evaluation already has, and it tells a reader which of your conclusions could survive a bad judge.

8. **The noise floor**: the share of rulings the same judge changes when it grades the same items twice. Section 5.6 is why: under Free it is nearly as large as a change of judge AI.

An evaluation can fill each item from records it already keeps, except item 8, which needs one repeat run. On our data, leaving the harness out of item 2 would have hidden a factor that changed 5.0 per cent of outcomes under Semi-Rigid, which we could not separate from noise without a repeat of each harness.

**This paper's own statement.** Where we did not record something, the table states it.

| Item | This paper |
|---|---|
| 1. Method | Rigid, Semi-Rigid, Free and the four-judge Estimative panel (section 5.2). Abstention is a counted "not sure" for the panel only |
| 2. Judge | Opus 5 and Gemini 3.1 Pro (preview). Snapshot IDs not recorded. Temperature and thinking at provider defaults, not recorded. Bare API: one request, output cap 8,192 tokens. Agent harness: a Claude Code session, no cap enforced, sampling set by the harness and not recorded (section 5.6) |
| 3. Rubric text and order | One prompt per method, shown with each submission in game order. The text is available on request |
| 4. Inter-judge agreement | Judge swaps in section 5.6, panel spread in section 5.7 |
| 5. Judge's share of variance | Section 5.4 |
| 6. Human audit | None. Planned (section 8) |
| 7. Load-bearing count | Section 5.9: 190, 76, 15 and 11 rulings |
| 8. Noise floor | Section 5.6: 2.9 to 6.1 per cent under Semi-Rigid, 13.5 to 15.9 per cent under Free |

### 6.3 What a practitioner should take from this

If you run exercises or evaluations, these are our main take-aways.

1. **Choose the method before you choose the judge.** The method decides how often each judge disagrees with itself (section 6.2, item 8). Under Free, a second run of the same judge changed nearly as many outcomes as a swap to another developer's judge AI: under our Free method, a different judge changed only about 2 to 4 points more.

2. **Do not audit a grading step by its pass rate alone.** Our panel changed 40 rulings and moved the pass count by 8. The pass rate stays still while the result changes underneath it.

3. **"We used a strong judge AI" is not a specification.** State the judge in full (section 6.2, item 2).

4. **When a ruling is built from several numbers, report their combined effect.** In our design the difficulty and the modifier combine into the implied odds of success. Our panel looks lenient on difficulty alone (5.91 against 9.29 for the rules), but it sits within 0.011 of the rules on implied odds.

5. **Count what your conclusions rest on before you argue about the judge.** The finding that every adjudicator's first run left intact rests on 190 rulings, and only one repeat moved it, in one of nine games. The finding that four of six adjudicators shifted, and every repeat moved, rests on 15. You can count this before you choose any judge.

---

## 7. Limitations, and what this cannot show

**There is no ground truth here, no human baseline, and therefore no accuracy claim anywhere in this paper.** This is the limitation to read before any other, because it bounds every sentence in section 5. We can say that an adjudicator moved a ruling. We cannot say the ruling was better. Nothing in this design contains a correct answer to compare against: the exercise had no correct answer to check against, which is the whole reason adjudication is the interesting problem, and we did not run a human adjudication panel over the same 378 submissions. So "the panel is less anchored" means less anchored, not more accurate. "Free flips 12.4 per cent of outcomes" means it disagrees with the rules on 12.4 per cent, and a reader who prefers the rules and a reader who prefers the panel can both point at that number. Where the paper recommends a method it is arguing from wargaming practice and from properties like reproducibility and contestability, never from measured correctness. The 40-ruling blind human audit in section 8 is the experiment that would change this, and it has not been run.

**The 378 rulings are not 378 independent draws, so the intervals are clustered by game.** They come from 9 games, 6 ticks, a small set of recurring actors and 3 evidence standards, with the same submission text recurring across seeds. Every interval in section 5.6 is a percentile cluster bootstrap at the **game** level, 9 clusters, 10,000 draws, resampling the same clusters across both cells of a contrast so a paired comparison stays paired. Nine clusters is thin. A percentile interval on nine clusters is a direction and a width, not three significant figures, and it is likely to undercover because the bootstrap cannot see between-game variation it has not observed. We also computed submission-level intervals, which are narrower, and we quote the game-level ones because they are the wider of the two.

Two consequences. Small differences tell us little: the gap between 4.8 and 5.0 per cent in section 5.5, and the 9.1 per cent rate read off 44 rulings in section 5.8, are descriptive and not effects. And most tables outside section 5.6 and Appendix D have no intervals, because the bootstrap was built for the contrasts the argument rests on rather than for every figure in the paper.

**We measured the noise floor in four of the experiment's cells.** The three seeds vary the d20 stream and the scripted run, not the judge, so the only within-cell replicates are the repeat legs in section 5.6: Gemini 3.1 Pro on the bare API and Opus 5 on the agent harness, under Semi-Rigid and Free. Bare Opus, the four-judge panel and Rigid were not repeated. Rigid needs no repeat, because a rule gives the same ruling twice. The repeats ran at provider defaults, not at a fixed low temperature, so they measure the noise of the settings in use, not the least noise these judges can give. Part of the 59 per cent residual in section 5.4's decomposition is this noise. The repeats show it changes 2.9 to 15.9 per cent of outcomes, depending on the method. The repeats ran six weeks after the first runs, on endpoints we did not pin, so each floor also includes any change in the served AI over that time.

**Section 5.9 is exploratory and uncorrected for multiplicity.** Four findings by six arms is a 24-cell table read for changes, alongside seven anchoring correlations, three band-width buckets, two variance tables and seven flip rates. We did not pre-specify which cells were primary and we apply no correction. Read it as hypothesis-generating.

**Sampling parameters were never set or recorded.** Section 1 of this paper faults evaluation papers for reporting "the AI's name, sometimes the temperature". We set no temperature at all, so every judge ran at its provider default, and we do not have those defaults logged. There is a second asymmetry in the same place: the bare API calls cap output at 8,192 tokens and the agent harness enforces no cap. Any comparison in section 5.6 between an agent-harness cell and a bare cell includes that difference too.

**All four judge AIs were served through one route.** Every bare API call went through one serving route, and the agent-harness calls ran in a Claude agent session. The panel's claim to independence is therefore a claim about four developers on one serving path, not about four independent serving paths. Shared infrastructure is a common-mode failure route that a panel is supposed to be robust to, and this design cannot rule it out.

**Judge AI identifiers are not pinned.** `google/gemini-3.1-pro-preview` is a moving preview endpoint, and `anthropic/claude-opus-5` is not a dated snapshot either. We give no snapshot identifiers or dates, so an exact rerun of these arms may not be possible.

**We audited our own engine.** This experiment is the Long Game Project re-adjudicating the Long Game Project. The mitigations are stated predictions, logs available on request, the planned replication on a second exercise, and the planned human audit. They are mitigations and not a substitute for someone else running it.

**Judge AI and harness are partly confounded**, per section 5.6. The Semi-Rigid and Free single-judge Opus arms ran on the agent harness, and every Gemini leg and every panel member ran bare. The bare Opus Semi-Rigid arm was added to break this, so two judge AI contrasts are clean (Semi-Rigid bare against bare, and Free with the lifted Opus panel member against the standalone Gemini arm) and two are not. Appendix D.3 labels every row. Any judge AI effect read off a row marked confounded is a statement about judge AI and harness together, and we do not rely on one.

**We measure rulings, not narratives.** Section 5.1 gives the reason: a game-level comparison is not identified, because a diverged trajectory changes the players' submissions as well as the adjudicator's rulings. So this paper measures how often the adjudicator changes a ruling, and section 5.9 says what that does to a published finding only to first order, with each trajectory held fixed. A full game-level answer needs the replays in section 8.

**Our judges see less than a human umpire.** Every judge, under every method, sees the scenario brief, the submission, the resource standing and its own earlier rulings in that game, but not the whole game state an NWC Free umpire would know (section 5.2). Free and Semi-Rigid get the same inputs, so the gap between them is not an information gap. A judge with the full game state may be more stable under both methods, and we have not tested that (section 8).

**Open is untested, and may be untestable here.** In NWC usage, Open adjudication is done in the presence of both player cells, with the results determined by the players (p. 28). Every arm here is Closed. Our engine has an Open variant, one step toward it: it returns the adjudicator's reasoning to the player for one rebuttal before the dice. A replay reads logged submissions, so the player cannot answer back. Open cannot be measured in this engine at all, and needs a live run. The legibility prediction is untested for that reason and not because it failed. The practitioner test, that players accept a defeat if they understand why they lost, is one we have not run.

**One exercise, one scenario, one domain, one engine.** Everything here is report 006. The replication on report 005, a different scenario in the same series, has not yet been run.

**The panel is four judge AIs, not four people.** Nothing here says a panel of four humans behaves this way, and the practice we drew it from was written about humans.

**Two judges is a thin crossed design.** Two methods by two judge AIs supports an interaction term and not much more. A third judge would let the decomposition give real weight.

**Our wargaming sources are practitioner literature.** Much of it is not peer-reviewed, and we present it as such rather than as established science.

---

## 8. Future work

In the order we would run them.

1. **Two repeat runs.** Repeat bare Opus, so both judges' floors are measured on the same harness. Run Free bare Opus as its own arm, to replace the panel member reused in section 5.6. They come first because the section 5.6 comparisons rest on them.
2. **Cross harness with judge AI.** We crossed harness with method inside Opus (section 5.6). Crossing it with judge AI needs an agent harness that serves the AI of more than one developer, which the current agent harness does not. Until that exists, the harness effect is established within Opus and not across developers.
3. **The 40-ruling human audit.** Two wargamers, blind to the AI rulings. This is the only test of which rulings are right. Every adjudicator is more generous than the players' own stated odds (section 5.5), and only a human audit can say whether that is calibration or leniency.
4. **Replicate on report 005**, a different scenario with 735 logged rulings.
5. **Agreement against discovery.** On a public rubric-judged clinical evaluation, vary rubric specificity across three levels and plant **legitimate** off-rubric behaviours: an honest refusal, a correct action the rubric author did not list, a safety escalation. Measure inter-judge agreement and planted-behaviour detection as specificity rises. Wargaming practice predicts agreement up and detection down. This is distinct from PReMISE, whose off-rubric behaviours are adversarial exploits, and the distinction between a legitimate off-rubric action and an exploit is the contribution.
6. **Run the four methods on a rubric-graded AI evaluation**, with a repeat run of each judge, to test whether section 5's result transfers from a game to ordinary AI grading.

7. **Judges that see the full game state**, under Semi-Rigid and Free, to test whether more context makes a judge more stable (section 7).
8. **Force the abstention** and find out whether a not-sure that a judge is made to consider conveys information or only compliance.
9. **A live Open run**, for the legibility prediction and the practitioner test that players accept a defeat they understand (section 7).

Shadow adjudication at a live human-refereed game is out of scope: the ethics approval on the exercise available to us is fixed and an amendment is not available.

---

## 9. Conclusion

AI evaluation now depends on judges, and research on them has asked which judge to use. Wargaming, where games have no correct answer either, asks what procedure the judge should follow. We mapped eleven of its practices onto AI evaluation failures, and tested one of them: the adjudication method, meaning who sets the difficulty and how far the judge may depart from a rulebook.

On 378 player actions with the dice held fixed, the method decided how often a judge disagreed with itself. When the judge could only adjust the rulebook's difficulty (Semi-Rigid), a second run of the same judge changed 2.9 to 6.1 per cent of outcomes. When it set the difficulty with no rulebook (Free), a second run changed 13.5 to 15.9 per cent. Both methods gave the judge the same inputs. A change of judge AI added about 3 percentage points to that under either method, a gap nine games cannot confirm. Of the two methods we ran twice, Semi-Rigid gave the more stable judges. On this evidence, choose the method of an AI-judged evaluation before the judge AI.

In this exercise, the findings that rested on fewer rulings moved more easily. A published finding that rests on 190 rulings survived the first run of every adjudicator. One that rests on 15 changed under most of the AI adjudicators we ran (section 5.9).

We cannot say which rulings were right. The exercise has no correct answer to check rulings against, and no human wargamers have yet ruled blind on the same actions. The evidence is also narrow: one exercise, nine games and two judge AIs. What we can say is how much the rulings move, and which conclusions rest on few enough rulings to move with them.

The fix is one paragraph in the methods. Publish an adjudication statement (section 6.2): the method, the judge AI with its harness and settings, a repeat run, and the count of rulings behind each conclusion. A reader can then see how much of a result is the judge's thumb on the scale.

---

## Declarations

**Use of artificial intelligence.** The author used Anthropic's Claude, through the Claude Code tool, between September and October 2026. The versions were Claude Opus 5 and Claude Opus 5.5. Working from the author's research question, design and replay records, the tool wrote the analysis scripts listed in Appendix C, drafted the manuscript text and checked bibliographic records against their sources. No client, participant or confidential material was given to the tool. The author directed each drafting round and reviewed and revised the text. Every reported number is generated by the scripts in Appendix C or counted from the replay records, not typed by hand. The AI systems studied in Section 5 are the experimental players and judges, and their use is described there. No AI tool is listed as an author. The author is accountable for the content.

**Competing interests.** The author is the founder of The Long Game Project, a consultancy that designs and runs wargames and tabletop exercises. The Long Game Project could benefit commercially from wider use of the adjudication methods discussed here. The author is a co-author of reference 10.

**Funding.** This work received no external funding.

**Ethics.** No human participants took part. Every logged submission came from a fixed script in an exercise refereed by AI.

**Data and code availability.** All 2,268 replay rulings and 2,194 repeat rulings with their reasoning, the run logs, the replay and analysis scripts in Appendix C, and their outputs are available from the author on request. The exercise engine is not released.

---

## References

Section 4 states which mapping-table claims have a page-level check.

Two notes on the sources. The *Wargame Pathologies* lead author is Weuve, not Perla. Downes-Martin also has a separate paper titled *Pathologies of War Gaming*, which is a different document from the CNA report and is not cited here.

---

## Appendix A: AusClinBench, as evidence and not experiment

AusClinBench is the author's separate benchmark of AI systems against current Australian clinical policy, with exact-match answer keys. Two of its items illustrate two rows of the mapping. Items are numbered in the benchmark's own scheme, such as item-123.

**The equipoise pairs.** A key that presupposes a directive answer marks an honest non-directive answer wrong. That is row 2 of the mapping in miniature: over-specification presupposing the answer.

**The item-123 case.** An AI key auditor sided with eight wrong AI systems against the primary source. That is row 5: consensus is not truth. Eight AI systems agreed, the auditor agreed with them, and all of them were wrong.

Neither is an experiment. Both are evidence.

## Appendix B: the aggregation rule is load-bearing

How the panel's four views combine changes the result, and the obvious rule is wrong.

The first version took the median of the difficulty column and the median of the modifier column separately. That rule fails on real rulings. Four judges on one submission returned difficulty 14 with modifier +3, difficulty 11 with +3, difficulty 17 with +9, and difficulty 12 with +6. The difficulty range is 6, which reads as violent disagreement. Their implied odds were 0.50, 0.65, 0.65 and 0.75, which is close agreement. The judge who said 17 and the judge who said 11 share the same view of the world, expressed with different bookkeeping.

Column-wise medians pair a difficulty with a modifier no judge gave. Across the full run the two rules disagree on 307 of 378 rulings, by 1.98 difficulty on average and by as much as 7.

So the panel collapses each member's pair into one implied probability, takes the median of those, and derives a bare difficulty back from it. That is what estimative probability means, and it generalises a rule we arrived at the hard way in the pilot: never report a difficulty without its modifier.

**Consequence for anyone building a panel.** Difficulty spread is not a disagreement measure. The spread of implied probabilities is.

## Appendix C: reproduction

All 54 legs and the 36 repeat legs, their rulings and their reasoning are on disk. Every number in this paper is regenerated by seven scripts and none is typed by hand:

```
python expa_analyse.py      # method against judge AI, per evidence standard
python expa_estimative.py   # the panel: anchoring, robustness, calibration, abstention
python expa_variance.py     # implied odds, variance, and the method-by-harness 2x2
python expa_loadbearing.py  # which of 006's findings survive, and what they rest on
python expa_intervals.py    # cluster bootstrap intervals on every contrast in 5.6
python expa_robustness.py   # per-game swaps, jackknife, pass counts
python expa_noise_floor.py  # the repeat legs: each judge against itself, and the swaps
        and the flips from the rules against that floor
```

The two figures are drawn from the same records by `build_figures.py`.

`expa_loadbearing.py` will not print results unless it first reproduces the engine's own recorded games from the logs: every resource at every tick of all nine scripted games, and all four finding flags. If the replayed effects ever stop matching, the script says so and reports nothing.

`expa_intervals.py` is deterministic. It seeds its resampler at 20260911, so the intervals in section 5.6 reproduce exactly rather than approximately. It makes no API calls.

Output collects into `EXPERIMENT-A-RESULTS.md`, which the scripts regenerate in full.

---

## Appendix D: section 5.6 in full

The tables and checks behind section 5.6, in the order the analysis ran. Every number is printed by `expa_intervals.py` or `expa_noise_floor.py` (Appendix C).

### D.1 The two harnesses

The two harnesses differ in these ways, and in no other setting we chose:

| | Bare API | Agent harness |
|---|---|---|
| Call | One Messages API request per ruling | The same request, answered by a Claude Code agent session behind a Messages-compatible endpoint |
| System prompt | The judge prompt only | The agent's own system prompt, then the judge prompt |
| Tools | None. The ruling is read from a JSON reply | The agent's tool loop is available. The ruling is read from the same JSON reply |
| Output cap | 8,192 tokens | None enforced |
| Temperature and reasoning | Provider defaults, not set by us | Set by the harness, not recorded |
| Mean output per Opus 5 ruling | About 3,000 tokens | About 3,500 tokens |

Both harnesses received the same judge prompt and the same submission.

That comparison is confounded with method, so we ran the missing cell, Semi-Rigid Opus 5 on the bare API over the same 378 submissions, and crossed the two factors. The prompt sizes recorded for bare API calls also differ between arms, for a reason we have not found, so we cannot claim every bare cell saw a prompt of identical size. All four cells below are Opus 5 on the same 378 submissions.

**One asymmetry.** Three of the four cells are standalone arms. The fourth, Free through the bare API, is the panel's Opus member lifted out of the Estimative runs. It is an independent adjudication call on the identical Free prompt, which is why we consider it usable, but it was not run as its own arm. Two consequences follow. The 14.6 per cent harness effect under Free compares a standalone agent-harness arm against a judge running inside the four-judge panel, so "harness" is not cleanly separated from "standalone against panel member". And this cell shares data with the panel result in section 5.7, so the two are not independent. The fix is one more arm, which we have not yet run.

### D.2 Harness crossed with method, Opus 5

| Cell | Mean difficulty | Mean modifier | Mean p(success) | Anchoring |
|---|---|---|---|---|
| Semi-Rigid, agent harness | 9.13 | 2.09 | 0.691 | -0.65 |
| Semi-Rigid, bare API | 9.42 | 2.94 | 0.714 | -0.60 |
| Free, agent harness | 11.68 | 7.24 | 0.745 | -0.41 |
| Free, bare API | 7.42 | 2.72 | 0.770 | -0.14 |

Now the outcome disagreements, each with one factor fixed:

| Contrast | Held fixed | Outcomes changed |
|---|---|---|
| Agent harness against bare API | Semi-Rigid | 19 of 378 (5.0%) |
| Agent harness against bare API | Free | 55 of 378 (14.6%) |
| Semi-Rigid against Free | agent harness | 52 of 378 (13.8%) |
| Semi-Rigid against Free | bare API | 40 of 378 (10.6%) |

The harness swap changes 5.0 per cent under Semi-Rigid and 14.6 per cent under Free. The noise floor in section 5.6 shows that neither clears Opus's own floor.

### D.3 Every adjudicator contrast, before the noise floor

Intervals are 95 per cent percentile cluster bootstraps over the nine logged games, 10,000 draws, resampling the same clusters across both cells of a contrast so the pairing survives. Section 7 says why the game and not the ruling is the cluster.

| Contrast | Held fixed | Harnesses | Outcomes changed | 95% CI |
|---|---|---|---|---|
| Method: Semi-Rigid against Free | Opus 5 | agent, agent | 52 of 378 (13.8%) | 9.5 to 18.0% |
| Method: Semi-Rigid against Free | Gemini 3.1 Pro | bare, bare | 48 of 378 (12.7%) | 9.5 to 16.1% |
| Method: Semi-Rigid against Free | Opus 5 | bare, bare (lifted) | 40 of 378 (10.6%) | 8.2 to 13.0% |
| Judge AI: Opus against Gemini | Semi-Rigid | bare, bare | 30 of 378 (7.9%) | 5.0 to 11.1% |

| Contrast | Held fixed | Harnesses | Outcomes changed | 95% CI |
|---|---|---|---|---|
| Judge AI: Opus against Gemini | Free | bare, bare (Opus lifted) | 66 of 378 (17.5%) | 13.5 to 20.6% |
| Judge AI: Opus against Gemini | Semi-Rigid | agent, bare (confounded) | 35 of 378 (9.3%) | - |
| Judge AI: Opus against Gemini | Free | agent, bare (confounded) | 69 of 378 (18.3%) | - |

A change of method moves 10.6 to 13.8 per cent of outcomes, and a clean change of judge AI 7.9 to 17.5 per cent. The noise floor in section 5.6 decides which of these are more than noise.

Stated as differences, with each difference computed inside the bootstrap draw so the correlation between the two rates is kept rather than ignored:

| Difference | Value | 95% CI (game clusters) |
|---|---|---|
| Method (agent) minus judge AI (Semi-Rigid) | +5.8 pp | -0.3 to +11.6 |
| Method (bare) minus judge AI (Semi-Rigid) | +2.6 pp | -1.9 to +6.6 |
| Method (agent) minus judge AI (Free) | -3.7 pp | -7.9 to +1.1 |
| Method (bare) minus judge AI (Free) | -6.9 pp | -10.1 to -3.7 |

Two of the four favour the method and two favour the judge AI. Only the Free difference excludes zero, and the noise floor in section 5.6 shows that most of that Free judge swap is the noise floor, so it is not evidence of a judge effect.

The two confounded rows are shown for completeness. Both pair an Opus leg on the agent harness against a Gemini leg on the bare API, so they mix a judge AI change with a harness change and cannot support a claim about either. The 18.3 per cent figure in particular is the largest number in the dataset and the least interpretable one, and it is not a judge AI effect.

### D.4 Judge AI swaps by method

Line the clean judge AI contrasts up by method:

| Method | Disagreement between judge AIs, harness matched | 95% CI |
|---|---|---|
| Semi-Rigid | 7.9% | 5.0 to 11.1% |
| Free | 17.5% | 13.5 to 20.6% |

Before the noise floor is removed, swapping the judge AI changes 2.2 times as many outcomes under Free as under Semi-Rigid (95 per cent CI on the ratio, 1.6 to 3.3).

Per game, Free changes more rulings than Semi-Rigid in 8 of 9 games, with one tie (one-sided sign test on the 8 untied games, $p = 0.004$).

| Game (evidence standard, seed) | Semi-Rigid swap | Free swap |
|---|---|---|
| behaviour, 1337 | 0 | 2 |
| behaviour, 20260904 | 6 | 8 |

| Game (evidence standard, seed) | Semi-Rigid swap | Free swap |
|---|---|---|
| behaviour, 7 | 5 | 9 |
| corroborated, 1337 | 6 | 6 |
| corroborated, 20260904 | 3 | 10 |
| corroborated, 7 | 4 | 7 |
| traces, 1337 | 2 | 7 |
| traces, 20260904 | 2 | 7 |
| traces, 7 | 2 | 10 |

Leaving out one game at a time, the ratio stays between 2.0 and 2.5. A jackknife interval with 8 degrees of freedom is 1.2 to 3.2, wider than the bootstrap, which can run narrow on nine clusters. Neither interval accounts for the judge's disagreement with itself. The noise floor in section 5.6 measures it, and it changes how this ratio should be read.

This matches the harness result in D.2, arrived at from the other direction. Semi-Rigid anchors the judge to the rules, and an anchored judge gives close to the same answer whichever harness it sits in. Free hands the judge the whole decision, and then the harness's system prompt and tool loop change the difficulty and modifier the judge sets. Under Free, the agent harness sets a higher difficulty and a larger modifier than the bare API (11.68 and 7.24 against 7.42 and 2.72). That is a level shift. We cannot separate from noise whether it changes which submissions pass. The two harnesses differ in more than one way: the agent harness adds a system prompt, a tool loop and no output cap. So this section measures a harness configuration, not one ingredient.

### D.5 Variance decomposition

This decomposition is a different design from section 5.4's. Section 5.4 crosses method with judge AI, both judges on the bare API, and finds method at 0.2 per cent. This one crosses method with harness inside Opus only, so its method share (4.2 per cent) is not comparable with 5.4's. It puts method above harness. Both are small:

| Source | Share of variance in implied odds |
|---|---|
| The submission itself (blocking) | 29.7% |
| Adjudication method | 4.2% |
| Harness | 0.8% |
| Method by harness | 0.0% |
| Residual | 65.3% |

Method sits above harness here, but both are small against a 65.3 per cent residual and we do not quote the ratio between them. More useful: the interaction reads 0.0 per cent while the disagreement table in D.2 shows a clear one. The interaction lives in which submissions flip, not in the average implied odds, so a decomposition on the mean cannot see it. Section 5.5 found the same: totals agree while individual rulings differ. So variance components reported on averages alone can hide an interaction.

**Two general claims the crossed design does not support.**

The harness does not move outcomes further than the judge AI in general. Under Semi-Rigid the harness changes 5.0 per cent of outcomes and the judge AI 7.9 per cent. Neither harness figure clears Opus's own floor (section 5.6).

The method does not move outcomes further than the judge AI in general either. It does on anchoring (section 5.3). On outcomes under Free, the judge AI swap exceeds the method change by 6.9 points, but the noise floor shows that swap is mostly the noise floor. The claim the data support is conditional: the method sets how often each judge disagrees with itself, and the judge AI adds about 3 points on top under either method (section 5.6).

### D.6 The noise-floor repeats

By mistake, each Opus repeat leg was started twice at once, so most submissions have two repeat rulings as well as the first run. We use all three: each submission's Opus floor is the mean of every pairwise disagreement it has. The three pairs give 2.4, 2.9 and 3.5 per cent under Semi-Rigid and 16.1, 17.0 and 14.3 per cent under Free. The two copies started from the same few early rulings and then saw some of each other's rulings as history, so they are not fully independent. One Opus reply gave a malformed modifier, and we used the other copy's ruling for that submission. One Semi-Rigid submission was never re-ruled, so that cell has 377.

Across every reasonable choice of Opus floor (pooled, the same-day pair only, the pairs against the first run, or one random copy) the judge-swap excess stays between 2.0 and 3.7 points. A reader may want the single-judge floors as well. The Semi-Rigid swap is 4.8 points above Opus's floor and 1.9 above Gemini's. The Free swap is 1.6 above Opus's and 4.0 above Gemini's. Neither single floor is the right comparison, because the swap includes both judges' noise.